\documentclass[12pt]{article}
     \usepackage[top=4cm, bottom=4cm, left=3cm, right=2.5cm]{geometry}
     \usepackage{setspace}
     \usepackage{footmisc}
     
     \makeatletter
     \newcommand\iraggedright{%
     \let\\\@centercr\@rightskip\@flushglue \rightskip\@rightskip
     \leftskip\z@skip}
     \makeatother
     \iraggedright
\usepackage{cite}
\usepackage{amsmath,amsthm, amscd, amssymb, amsfonts}
\usepackage{appendix}
\usepackage{enumerate}
\usepackage{chicago}
\begin{document}
\title{On the polemic assessment of what Bell did\\
{\small
Facultad de Ciencias Exactas y Naturales,Universidad Nacional de Asunci\'{o}n,Ruta Mcal. J. F. Estigarribia, Km 11 Campus de la UNA, San Lorenzo-Paraguay
}}
\author{\small{Justo Pastor Lambare\footnote{email: jupalam@gmail.com}}}
\date{}
\maketitle
\begin{abstract}
Despite their Nobel Prize-winning empirical falsification, the interpretation of Bell's inequality remains a subject of controversy.
This article discusses and attempts to clarify the reasons John S. Bell and A.  Einstein claimed that quantum entanglement implies puzzling nonlocal correlations that Einstein famously termed ``spooky action at a distance.''
The issue remains notoriously controversial and has roughly divided the scientific community into localists and nonlocalists.
Without taking a stance on either side in the long-standing, polarized debate, we examine Bell's actual argument and highlight how his reasoning differs from the current orthodoxy.
\end{abstract}

\newpage
%
%
%
\section{Introduction}\label{sec:Intro}
The intention of this work is neither to challenge the current dominant interpretations nor to solve the long-standing controversy on the nonlocality problem, but to expose Bell's true ideas by analyzing his writings with an objective and undogmatic attitude, disentangling what Bell's argument indeed was and highlighting the subtle yet important conceptual differences with the prevailing orthodoxy.

Although we subscribe to what we sustain was Bell's reasoning, we refrain from claiming that Bell's views on quantum mechanics are the correct ones.
We limit ourselves to examining the existing textual evidence and accordingly expose his reasoning on the quantum nonlocality problem, explaining how ongoing widespread arguments differ from Bell's approach.

Regarding quantum mechanics, the scientific community's interpretation is generally divided into localists and nonlocalists.
However, unlike Bell's approach, both groups invariably base their arguments on the precept that the proof of quantum nonlocality relies on the violation of the Bell inequality.

A characteristic exposition of this dichotomy is well depicted by an exchange between Tim Maudlin, with his article ``What Bell did'' which suggested the title of this paper, and Reinhard Werner's comments, exemplifying the innumerable articles claiming quantum nonlocality based on the Bell inequality, on the one hand, and its rejection for the same reason, on the other \cite{pMau14,pWer14,pMau14a,pWer14a}.

In section \ref{sec:HRofBWN} we present evidence supporting that Bell did not claim his inequality is a proof of quantum nonlocality.
The section has a historical character introducing a chronological review of Bell's main papers on nonlocality.
It shows that, in his most lucid writings, Bell formulated his inequality only after explicitly establishing the nonlocal character of quantum mechanics, either by previously assuming it \cite{pBel64} or by directly proving it \cite{pBSHC85,bBel90}.

Section \ref{sec:BPQN} delves into Bell's actual quantum nonlocality argument, which is not based on EPR but on local causality, a concept he introduced in 1975, more than ten years after the publication of his celebrated 1964 paper.
We highlight a formal application of local causality to quantum mechanics (sec. \ref{sssec:QNL}), and identify the origin of the longstanding dispute between localists and nonlocalists (sec. \ref{ssubsec:Tgoec}).

We advance the three principal points that clash with the usual prevalent views, but that, according to the existing evidence,  were sustained by Bell:
\begin{itemize}
\item Orthodox operational quantum mechanics is nonlocal because it lacks local common causes to explain perfect correlations, not because it violates the ``locality inequality.''
\item Since quantum mechanics is nonlocal in the above sense, the only way to avoid nature's nonlocality is to complete quantum mechanics with local common causes $\lambda$ that would circumvent the eventual existence of ``action at a distance.''
    The admissibility of this last point is what the Bell theorem is all about, not quantum nonlocality per se.
\item The $\lambda$ variables, understood as common causes, are not necessarily ``elements of physical reality'' in the EPR sense \cite{pEPR35}. Indeed, although Bell often referenced the EPR paradox, he never resorted to the concept of elements of physical reality.
Moreover, he explicitly mentioned that the $\lambda$ variables could well include the quantum state vectors (cf. sec. \ref{sssec:QNL} and appendix \ref{app:CCM}).
Thus, according to Bell’s reasoning, the $\lambda$ common causes variables are fairly general parameters. 
They are wholly compatible with quantum nonrealism,\footnote{By realism in this paper, we mean the usual connotation given in physics, i.e., the view according to which physical magnitudes are assumed to exist and have definitive values whether or not they are observed. } do not necessarily disrupt quantum superposition, and are not conditioned to be hypothetical values that preexist before actual measurements.
\end{itemize}
The subtle difference between the first two points is relevant to appreciate the consistency of Bell's argument. Establishing quantum nonlocality is prior and independent of the formulation of the ``locality inequality,'' i.e., he made a clear distinction between the problems of quantum nonlocality and quantum completion and, contrary to the current orthodoxy, did not conflate them.

The third point is relevant to Bell's proof that orthodox operational quantum mechanics violates the local causality condition; a concept he only introduced much later, in the mid-1970s, to explicate quantum nonlocality.
The proof is independent of the EPR reasoning and, of course, the ``locality inequality.''

Bell's local causality concept aligns with the \textit{Principle of Common Causes}, which was independently and more completely developed by Reichenbach \cite{bRei56} to study the problem of causal structures and probabilistic correlations.
According to Cavalcanti and Lal \cite{pCav14}:
\begin{quote}
That Reichenbach did not arrive at Bell's theorem even though he considered both common causes and causal anomalies in quantum theory goes to show how subtle and deep was Bell's insight.
\end{quote}
\section{Chronological Review of Bell's Works on Nonlocality}\label{sec:HRofBWN}
We present a brief review of Bell's main papers on nonlocality. The key point is that nowhere in those papers can we find an explicit statement declaring that quantum mechanics' nonlocal character is a direct consequence of the violations of his inequality.
On the contrary, as we shall see, he always avoided stating such an inference.

On the other hand, he explicitly proved quantum mechanics violates local causality before formulating any inequality at least on two occasions \cite{pBSHC85,bBel90}.
Therefore, contrary to the usual folklore, the objective textual evidence indicates that Bell did not consider his inequality as a direct proof of quantum nonlocality but only of the impossibility of a non-conspiratorial local completion.\footnote{Non-conspiratorial means that the statistical independence condition is assumed (cf. \ref{ssec:SI}).}
\subsection{The 1964 Bell Theorem}\label{ssec:1964}
According to the prevalent view, Bell formulated his inequality to prove that quantum mechanics is not a local theory, presenting the Bell theorem as a quantum nonlocality theorem.
But an attentive reading of Bell's 1964 argument reveals two crucial facts that are widely overlooked and prove otherwise \cite{pBel64}:
\begin{enumerate}[a)]
\item Bell already considered quantum mechanics as nonlocal from the beginning, i.e., before formulating his inequality. Indeed, in the third line of the introduction, he wrote: \emph{``These additional variables were to restore to the theory causality and locality.''}
    That is, the addition of hidden variables to the theory was supposed to modify it to recover locality, instead of proving its nonlocality.
\item In correspondence with the above-quoted sentence, Bell starts the conclusion section by saying: \emph{``In a theory in which parameters are added to quantum mechanics....''}; so, clearly, he was not inferring properties of quantum mechanics, but only of a modified theory in which parameters are added.
\end{enumerate}
Thus, Bell's stunning conclusion is not that quantum mechanics is nonlocal, which he, as Einstein, took for granted, but that we cannot fix its nonlocality by completing it, something that would have surely disappointed Einstein.

Bell resumed the argument where EPR has left it: if local, quantum mechanics must be incomplete.
Since the orthodox interpretation asserts quantum mechanics is complete, then, according to EPR, it must be nonlocal (cf. sec. \ref{ssec:TEPRP}).

The previous inference was implicit in his expression referenced in a), asserting that additional variables were necessary to restore locality.
Then, applying an EPR-like reasoning, Bell derived a deterministic hidden variable model and proved, through his inequality, the untenability of a local completion.\footnote{Recently, Michael Hall claimed that deterministic hidden variables cannot be derived from Bell's 1964 hypotheses. However, as Hall himself recognizes, the issue is not relevant because later generalizations of the inequality did not assume perfect correlations; besides, stochastic Bell inequalities were formulated \cite{pHal24}.}

Thus, accepting the EPR reasoning, as Bell did, the inequality is unnecessary to prove quantum nonlocality, so claiming that Bell's inequality proves it (which Bell did not) would only be circular reasoning.

The previous step of accepting the EPR reasoning as proof of quantum nonlocality is crucial for maintaining the consistency and relevance of Bell's 1964 argument.
Otherwise, after formulating his inequality, he would have ended up with an already local quantum mechanics theory that just cannot be locally extended with additional variables, i.e., it would be possible to think that the non-local effect was a consequence of the addition of hidden variables that were not present at the beginning; by the way, the standard localists' argument.

Bell's theorem is a simple mathematical theorem that should be free of any polemic if we strictly follow Bell's rationale, namely, that a local completion is untenable.
Probably, that was what Richard Feynman meant when he said that Bell's theorem \cite{pWhi84},
\begin{quote}
``It is not an important theorem. It is simply a statement of something we know is true -- a mathematical proof of it.''
\end{quote}
However, we disagree with Feynman on the unimportance of the Bell theorem since it was a significant advance in the Bohr--Einstein debate that, by 1964, remained stagnant for almost thirty years.
\subsection{Bell's theorem after 1964}\label{sec:a1964}
Bell's arguments evolved over the years.
In later papers, although he occasionally mentioned EPR, he mostly abandoned the EPR reasoning. 
Whenever explicitly proving quantum nonlocality, he always used the concept of local causality, which he conceived to be independent of EPR and determinism.

As in his 1964 paper, in the articles he published afterward, we could not find any convincing evidence suggesting that he considered his inequality to be a proof of quantum nonlocality. 
However, there is clear evidence to the contrary.

Although he generalized the ``locality inequality'' to a stochastic version, there is sufficient evidence indicating how he continued interpreting his inequality as proving only the impossibility of a local completion.
Indeed, at least on two occasions \cite{pBSHC85,bBel90}, he unambiguously concluded that quantum mechanics is not locally causal before introducing hidden variables and formulating any inequality.

On other occasions where he might not have been so explicit, it was probably because he assumed that quantum nonlocality was an accepted trait of quantum weirdness, so he moved more directly to the local completion problem, unfortunately leaving some room for misinterpreting his approach.
However, such misinterpretation becomes evident when considering the totality of his works in a consistent whole.

To prove our previous assertion, next, we chronologically examine Bell's main papers discussing nonlocality after 1964.
\subsubsection{Introduction to the hidden variable question}\label{ssec:Ithvp}
In 1971, he wrote the paper ``Introduction to the hidden-variable question'' \cite{bBel71}.
Here, Bell did not address the question of quantum nonlocality but, as the title explicitly reveals, investigated hidden variable theories using the de Broglie--Bohm theory as a paradigmatic example, stressing its conspicuous nonlocal character as the ``difficulty.''

His paper highlights the fact that de Broglie--Bohm theory is not nonlocal incidentally, but that any extension of quantum mechanics must necessarily be nonlocal.
Bell proved that quantum mechanics violates his inequality, concluding:
\begin{quote}
``Thus the quantum-mechanical result cannot be reproduced by a hidden-variable theory which is local in the way described.''
\end{quote}
Note that the adjective ``local'' in his expression refers to the hidden-variable theory, so irrespective of whether he believed that quantum mechanics is nonlocal, he did not say his inequality proved it.
Any assertion in that direction would be unjustified and contrary to what he indeed used to explain elsewhere when explicitly proving quantum nonlocality.
\subsubsection{The theory of local beables}\label{ssec:Ttolb}
This article appeared in 1975.\footnote{Bell's work is reproduced in \cite{pBSHC85}.}
Here, Bell abandoned the EPR reasoning and set out to prove quantum nonlocality explicitly by introducing the concept of local causality.
He argued that quantum mechanics violates this form of locality in section 3 without mentioning any inequality.
He starts that section by asserting:
\begin{quote}
``Ordinary quantum mechanics, even the relativistic quantum field theory, is not locally causal in the sense of (2).''
\end{quote}
(2) above refers to the local causality constraint, a concept he developed in the previous section.
Then, he presented his argument on quantum nonlocality.
In the same section, but only after establishing the nonlocal character of quantum mechanics, Bell explored the problem of adding hidden variables.

Then in section 4, ``Locality inequality,'' he derived a stochastic Bell--CHSH inequality.
Finally, in section 5, he established the impossibility of a local completion by proving that quantum mechanics violates his inequality, concluding:
\begin{quote}
``So quantum mechanics is not embeddable in a locally causal theory as formulated above.''
\end{quote}
That is different from concluding, ``So quantum mechanics is not a local theory'' unless we force the interpretation by dismissing his procedure.
Otherwise, why would he bother to first prove quantum mechanics violates local causality two sections before without using any inequality or hidden variables?
We guess because he agreed with Stapp about the logical loophole of concluding quantum nonlocality directly from his inequality \cite{pSta12}:
\begin{quote}
``Thus whatever is proved [with the Bell inequality] is not a feature of quantum mechanics, but is a property of a theory that tries to combine quantum theory with quasi-classical features that go beyond what is entailed by quantum theory itself. One cannot logically prove properties of a system by establishing, instead, properties of a system modified by adding properties alien to the original system.''
\end{quote}
Above, ``properties alien to the original system'' rigorously mean variables that do not legitimately pertain to quantum mechanics.
Although some have observed that the hidden variables can include the quantum state  \cite{pNor11,pGis12,pLau18}, the problem persists with the other ``additional variables.''
As we observe in the section \ref{ssubsec:Tgoec}, the Bell inequality cannot be formulated without additional variables foreign to quantum mechanics, notwithstanding that one of those variables may include the quantum state as indicated by Bell himself \cite{pBel81}.
\subsubsection{Bertlmann's socks}\label{ssec:Bs}
In 1981 Bell wrote his celebrated paper ``Bertlmann's socks and the nature of reality'' \cite{pBel81}.
On this occasion, Bell did not explicitly prove quantum nonlocality before formulating the inequality.
He based his arguments on EPR, but he also mentioned common causes to explain local correlations.

This is one of the papers where he left ambiguous whether his inequality violation should be interpreted as proof of quantum nonlocality.
Can we assume that Bell changed his mind about the meaning of his inequality?
We do not think so because, in his last paper (cf. \ref{ssec:Lnc}), he returned to his previous formulations, i.e., either assuming \cite{pBel64} or proving \cite{pBSHC85} quantum nonlocality without introducing hidden variables or mentioning any inequality.

In ``Bertlmann' socks'', Bell chose intuition and ease of interpretation over logical rigor.
In Bell's own words, this paper was one of those that \cite{bBel04}:
\begin{quote}
``...are nontechnical introductions to the subject. They are meant to be intelligible to nonphysicists.''
\end{quote}
That is why he spent great effort explaining the difference between quantum and classical entanglement through naive analogies, such as those of the socks of Mr. Bertlmann.
It is rather surprising that some scientists still believe that quantum entanglement admits such a trivial explanation (cf. sec. \ref{app_sssec:Aace}).
\subsubsection{La nouvelle cuisine}\label{ssec:Lnc}
This is Bell's last paper which appeared in 1990 \cite{bBel90}.
Here again, Bell's view of his inequality and quantum nonlocality is crystal clear.

This time his approach differs from his 1975 formulation in ``The theory of local beables'' in that he proves that quantum mechanics violates both nonlocality criteria, EPR and also local causality.
However, what remains unchanged is that he proved quantum nonlocality in section 8 before formulating his inequality.

Only in section 10, he formulates his inequality and mentions that quantum mechanics violates it, concluding that:
\begin{quote}
``Quantum mechanics cannot be embedded in a locally causal theory''
\end{quote}
Although some interpret the above expression as proving that nature itself, including quantum mechanics, is nonlocal, the order in which he presents his argument is unambiguous.
First, establishing quantum nonlocality without any inequality, and then, in a separate section, inferring his conclusion through his inequality.

So, at least respecting Bell's reasoning, we cannot conclude that his inequality alone leads to inferring that nature itself is nonlocal unless we dismiss the first part of his argument, where he independently, without an inequality, proved that quantum mechanics is nonlocal.
\subsubsection{Further writings}
In 1975, an editor asked Bell to write a response letter \cite{pBel75a}. He started the paper by expressing:
\begin{quote}
The editor has asked me to reply to a paper, by G. Lochak, refuting a theorem of mine on hidden variables.
\end{quote}
Despite the paper's title being \textit{Locality in quantum mechanics: reply to critics}, it is remarkable that he referred to his theorem as ``a theorem of mine on hidden variables'' instead of ``a theorem of mine on quantum nonlocality.''
Again, although both subjects are intertwined, he carefully maintained a distinction.

In 1977 Bell wrote \textit{Free variables and local causality} \cite{pBel77}. He started the paper by writing:
\begin{quote}
It has been argued that quantum mechanics is not locally causal and cannot be embedded in a locally causal theory
\end{quote}
Once more, his expression openly shows his reasoning.
He considered the statement ``quantum mechanics is not locally causal'' to be before, different from, and not a consequence of, the assertion ``cannot be embedded in a locally causal theory'' which was the conclusion he used to draw from the violation of his inequality as an additional characteristic of quantum mechanics besides nonlocality, as he explicitly did it in Refs. \cite{pBSHC85,bBel90} and we explain in sections \ref{ssec:Ttolb} and \ref{ssec:Lnc}.
\section{Bell's Proof of Quantum Nonlocality}\label{sec:BPQN}
In this section, we analyze Bell's explicit argument on the nonlocal character of quantum mechanics.
In his 1975 paper, ``The theory of local beables'' \cite{pBSHC85}, Bell gave an explicit argument for quantum nonlocality for the first time.
This paper has four outstanding characteristics that were missing in 1964:
\begin{itemize}
\item A formal definition of locality that is directly applicable to quantum mechanics. He called it \emph{Local Causality} (LC).
\item An argument showing that quantum mechanics violates LC and hence is not locally causal. Bell presented his nonlocality argument before formulating his inequality, therefore, despite the usual claim, he did not consider the former a consequence of the latter.
\item A justification for assuming \emph{statistical independence} (SI) in his hidden variable model. In 1964, SI was an \emph{ad hoc} implicit assumption.
\item An absence of any reference to the EPR paper.
\end{itemize}
Next, we briefly address each of these characteristics.
\subsection{Local Causality and Quantum Nonlocality}\label{ssec:LCandQN}
Firstly, we briefly review the meaning of LC.
Unfortunately, as observed by Norsen \cite{pNor11}, LC is a little-known concept, and we would add that, if known, it is often misunderstood.
For this reason, appendix \ref{app:LC} contains a more detailed explanation for those unfamiliar with it.
Secondly, like Bell, we prove that quantum mechanics violates LC without the Bell inequality.
Thirdly, we analyze the origin of the argumentative conflict between localists and nonlocalists.
\subsubsection{Local causality}\label{sssec:LC}
The concept of LC was conceived to be directly applicable to not deterministic theories like quantum mechanics.
Here it suffices to say that when distant experiments are independently performed, to exclude nonlocal effects, the existence of correlations must have a local common cause explanation.

More concretely, in the case of a Bell-type experiment, if $P(A,B\mid a,b)$ is the joint probability for Alice and Bob finding results $A$ and $B$ when their experimental settings are $a$ and $b$ respectively, we in general have that,
\begin{eqnarray}\label{eq:cr} 
P(A,B\mid a,b) &\neq& P(A\mid a) P(B\mid b)
\end{eqnarray}
(\ref{eq:cr}) means the results of the experiment are correlated.
However if correlations are to be explained locally, after including all relevant factors represented by $\lambda$,\footnote{In case many common causes are required, $\lambda$ represents a vector variable.} we must have,
\begin{eqnarray}\label{eq:ucr} 
P(A,B\mid a,b,\lambda) &=& P(A\mid a,\lambda) P(B\mid b,\lambda)
\end{eqnarray}
(\ref{eq:ucr}) means that after all factors (known and unknown) are included, whatever Bob decides to do in his distant laboratory cannot influence Alice's local measurements; and vice versa.
Therefore, (\ref{eq:ucr}) must be satisfied by locally causal theories.
Note that $\lambda$ represents local common causes in general, not necessarily ``preexisting values'' and can well include the quantum state (cf. appendix \ref{app:CCM}).

\subsubsection{Quantum nonlocality}\label{sssec:QNL}
After defining local causality, Bell gave an argument explaining why, if considered complete, quantum mechanics violates it.
The conditional ``if considered complete'' is crucial to understand what he meant by quantum nonlocality, namely, that the quantum state alone cannot locally explain the predicted perfect correlations which does not mean that nature itself is nonlocal unless an alternative hypothetical ``more complete'' theory is impossible.

The Bell theorem deals with the possibility of certain kinds of ``more complete'' theories, not directly with quantum nonlocality as understood above.

In \cite{pBSHC85}, Bell gave a qualitative description of why a locally causal explanation is absent and concluded that in quantum mechanics ``We simply do not have (2)'', where (2) in his paper is LC.

It is relevant to realize that by locally causal explanation he meant any explanation, not only ``real'' explanations in the sense of preexisting entities.
His argument is similar to the one given by Einstein in 1927.\footnote{Einstein's argument is explained by Laudisa \cite{pLau19} and also by Harrigan and Spekkens \cite{pHar10}. It is a genuine nonlocality argument that does not contain speculations about the infamous ``elements of physical reality.''}
Both Einstein and Bell were concerned with the existence of ``actions'' or ``influences'' at a distance. They were not concerned with the need for preexisting entities, determinism, or the denial of quantum superposition that are commonly associated with the notion of realism. 
Those were subsidiary and irrelevant issues regarding their true concerns about nonlocality.\footnote{Those issues were obfuscated in the EPR paper by the ``elements of physical reality'' concept, which neither Einstein nor Bell used in their arguments \cite{pHow85}.}

We can recast Bell's and Einstein's arguments in more formal terms through the mathematical formulation of local causality.
The crucial point is that (\ref{eq:ucr}), unlike the EPR reasoning, avoids a deterministic formulation or any ``classical'' prejudice and is directly applicable to quantum mechanics.
If quantum mechanics is complete and local, the locally causal explanation of its correlations must lie within the quantum state.
In our case
\begin{equation}\label{eq:ss}
\mid \psi\rangle=\frac{1}{\sqrt{2}}(\mid +\rangle \otimes \mid -\rangle-\mid -\rangle \otimes \mid +\rangle)
\end{equation}
Thus, if locally causal, ordinary quantum mechanics must satisfy (\ref{eq:ucr}) when the local common cause is equated to the quantum state,
\begin{equation}\label{eq:cciqs}
\lambda=\mid \psi\rangle
\end{equation}
However, choosing $a=b$, $A=1$, and $B=-1$ in (\ref{eq:ucr}), an elementary  quantum mechanical calculation gives
{\small
\begin{equation}\label{eq:sslc}
\underbrace{P(1,-1\mid a,b,\mid\psi\rangle)}_{1/2} \neq \underbrace{P(1\mid a,\mid \psi\rangle)P(-1\mid b,\mid \psi\rangle)}_{1/2\,*\,1/2\,=\,1/4}
\end{equation}
}
which does not comply with (\ref{eq:ucr}) since in (\ref{eq:sslc}) $1/2\neq 1/4$, meaning that, although the quantum state lies within the past light cone of the measuring events, it cannot screen off events on one side from spacelike separated events on the other far away side.

Given that (\ref{eq:sslc}) is not widely known as a proof of quantum nonlocality, appendix \ref{app:LC} further explains the local causality concept, and appendix \ref{app_sssec:Aace} clarifies a common misunderstanding claiming to refute (\ref{eq:sslc}) via a counterexample.

Also, some reject (\ref{eq:cciqs}) as incorrect because it would somehow conflict with quantum ``nonrealism.''
However, according to Bell, ``...nothing is said about the locality, or even localizability, of the variable $\lambda$. These variables could well include, for example, quantum mechanical state vectors, which have no particular localization in ordinary space-time...'' \cite{pBel81}.
Thus, in Bell's view, and lining up with Reichenbach's Common Cause Principle, it is unjustified to restrict $\lambda$ to be an ``element of physical reality'' in the EPR sense (cf. appendix \ref{app:CCM}).

So, according to Bell's argument, quantum mechanics violates local causality without any subterfuge of realism sneaking in.
The violation of (\ref{eq:ucr}) formally means that we can not find a locally causal explanation for the perfect correlations within quantum mechanics, be it ``real'' or ``unreal'' in the quantum mechanical sense.
If such local explanation exists, it must be found elsewhere.
\subsubsection{The gist of the endemic controversy}\label{ssubsec:Tgoec}
Without entering obscure metaphysical digressions about realism or ``classicality'', the concrete reason why arguing quantum nonlocality from the Bell inequality is unconvincing is that proving Bell-type inequalities requires writing joint probabilities as,
{\small
\begin{equation}\label{eq:ineqprob}
P(A,B\mid a,b)=\int P(A\mid a,\lambda)P(B\mid b,\lambda)P(\lambda) d\lambda
\end{equation}
}
which is impossible without going beyond quantum mechanics since (\ref{eq:sslc}) proves that
{\small
\begin{equation}\label{eq:nucr} 
P(A,B\mid a,b,\mid\psi\rangle) \neq P(A\mid a,\mid\psi\rangle) P(B\mid b,\mid\psi\rangle)
\end{equation}
}
Thus, we are confronted with the following facts:
\begin{enumerate}[1)]
\item The proof of any property based on (\ref{eq:ineqprob}) is not a property that can be unambiguously ascribed to quantum mechanics.
\item (\ref{eq:nucr}) is proved independently of (\ref{eq:ineqprob}), so the Bell inequality is not necessary to prove that orthodox operational quantum mechanics' predictions indeed violate the local causality condition.
\end{enumerate}
For some strange reason, those who argue quantum mechanics is nonlocal dismiss (\ref{eq:nucr}) as proof despite, in some cases, being well aware of its existence.
They prefer to turn to the Bell inequality as their sole argument, conflating the issues of quantum nonlocality with the possibility of a local completion \cite{pNor11}, something that, as we have seen, Bell was careful to distinguish.

On the other hand, those claiming that quantum mechanics is local rely on one of the following strategies:
\begin{itemize}
  \item recursing to point 1) above, conveniently blocking the ``usual'' quantum nonlocality proof.
  \item invoking realism or the ``elements of physical reality'' idea that is universally attached to the EPR reasoning. 
\end{itemize}
The first point above is arguably well justified for rejecting quantum nonlocality.
However, the second one, although very popular, lacks rational support because Bell's local causality was conceived independently of any notion of determinism or preexisting values and cannot be consistently applied to this case.
\subsection{Statistical Independence}\label{ssec:SI}
As we mentioned above, in Bell's 1964 paper, he implicitly assumed the hidden variables distribution function\footnote{Note that the distribution function of the $\lambda$ common causes is irrelevant for the definition of local causality. $P(\lambda)$ is necessary only to derive the Bell inequality.} $P(\lambda)$ was not conditional on the experimental settings $a$ and $b$.
\begin{equation}\label{eq:si}
   P(\lambda\mid a,b)=P(\lambda)
\end{equation}
We can justify (\ref{eq:si}) by requiring the experimental settings to be independent of the same common factors $\lambda$ affecting the results
\begin{equation}\label{eq:is}
   P(a,b\mid \lambda)=P(a,b)
\end{equation}
According to Bayes theorem we have
\begin{equation}\label{eq:bayes}
   P(a,b\mid \lambda)P(\lambda)=P(\lambda\mid a,b)P(a,b)
\end{equation}
Then from (\ref{eq:is}) and (\ref{eq:bayes}) we get (\ref{eq:si}).
The ansatz (\ref{eq:is}) seems to be a reasonable assumption justifying (\ref{eq:si}).

Thus, (\ref{eq:is}) and (\ref{eq:si}) are equivalent and are known as \emph{statistical independence}, \emph{measurement independence}, \emph{freedom}, or \emph{no-conspiracy}.
\subsection{The EPR Paper}\label{ssec:TEPRP}
Although in 1964, Bell conceived his article as a continuation of the EPR paper (assuming quantum nonlocality to investigate the possibility of a local completion), it should not be surprising that in his 1975 paper ``The theory of local beables'' \cite{pBSHC85}, Bell did not reference the EPR manuscript. 
This time, his intention was to provide an argument independently of EPR and determinism.

Although the EPR reasoning was intended to be a proof for quantum incompleteness according to the implication,
\begin{equation}
Locality \rightarrow Incompleteness
\end{equation}
the same reasoning can be conceived as a quantum nonlocality proof by the contrapositive implication,
\begin{equation}\label{li:cinl}
\begin{array}{lcl}
\neg Incompleteness &\rightarrow& \neg Locality\\
Completeness &\rightarrow& Nonlocality
\end{array}
\end{equation}
Perhaps motivated by the orthodox criticism against the EPR reasoning, unlike in his 1964 approach, the above (\ref{li:cinl}) quantum nonlocality proof was not implicitly assumed in his 1975 paper, which is why he introduced the concept of local causality to provide an independent proof.
The proof was directly applicable to non-deterministic theories such as quantum mechanics, eliminating the realism tenet that, to Einstein's dislike, somehow obscured the EPR argument \cite{pHow85}.

However, it is also true that Bell and Einstein considered the basic EPR reasoning to be correct, notwithstanding the unnecessary metaphysics surrounding the popular ``elements of physical reality'' concept, which somehow blurred the argument and was never used by either Bell or Einstein in their personal accounts.
\section{Conclusions}\label{sec:Conclu}
The purpose of this piece is not to challenge the established orthodoxy, but to analyze what Bell’s true argument was according to an objective analysis as it stands in the written evidence he left us.

According to Bell’s approach, quantum nonlocality and quantum incompleteness, although somehow related, are logically different issues.


Unlike the conventional interpretation, a careful analysis of Bell's papers reveals that he considered his theorem to be, as he once put it,  ``... a theorem of mine about hidden variables'' \cite{pBel75a}, rather than a theorem about quantum nonlocality which, according to him, demanded an independent proof.

Logical rigor suggests following Bell's approach; otherwise, an argumentative loophole may arise: if we do not first prove that orthodox quantum mechanics is nonlocal, then the fact that a different modified theory is nonlocal does not necessarily imply the nonlocality of the first one.

Thence, the omission of the previous first step renders localists with a convenient counterargument for claiming that the Bell inequality only proves a ``distorted'' non-quantum theory is nonlocal, not quantum mechanics itself.
It is rather ironic that Bell probably would have agreed with them on that specific point.

As per Bell's reasoning, the persistent controversy over whether his theorem disproves realism or quantum locality would be a misleading and ill-posed dilemma.

Understanding Bell's and Einstein's reasoning does not mean accepting quantum nonlocality; however, it reveals that widely spread and endless disputes are probably ill-formulated, hindering the advancement to deeper insights and obfuscating the general perception of the problem.

A rational alternative for arguing quantum locality could be to reject local causality as a valid criterion of locality and postulate that quantum mechanics is local. 
Given its no-signaling property, this is logically and empirically admissible, but a seldom, perhaps never, explicitly admitted path. 
Instead, obscure and unconvincing metaphysical speculations about realism are rehearsed, which are logically disconnected from the nonlocality problem, thereby exacerbating the existing controversy \cite{pNor07,pGis12}.

A probably more drastic alternative is to embrace superdeterminism and consider quantum mechanics as an emergent theory \cite{pHoo21}.
\newpage
\section{APPENDIX}
\subsection{Local Causality}\label{app:LC}
Here we present a more detailed analysis of the local causality concept and give an example of a common misunderstanding.
We refer the reader to the excellent exposition of local causality by Norsen \cite{pNor11}.
\subsection{The LC concept}
Bell's definition of LC is a formalization of the idea that, according to relativity theory, interactions can happen only at a finite speed.
It means that causes cannot have an instantaneous effect on distant events.
He formulated LC so that it can be applied to nondeterministic theories like quantum mechanics.
It is a locality argument that avoids a purportedly classical EPR-like reasoning.
A concept directly applicable to orthodox quantum mechanics without distorting its nature.

For the particular case that concerns us, i.e., the singlet state correlations in a Bell-type experiment, LC takes the following form.
Let $P(A,B\mid a,b)$ be the probability of a joint measurement giving the results $A,B\in\{-1,+1\}$ conditional on the respective measurements directions $a,b$. The laws of probabilities require
\begin{equation}\label{eq:lawofp}
P(A,B\mid a,b)=P(A\mid B,a,b)P(B\mid a,b)
\end{equation}
So far, it is just about probabilities.
Let us now add some physics and assume that both observers, Alice and Bob, choose their measurement directions at the last moment so that both measurements are spacelike separated events.
Then LC requires that neither the results $A,B$ nor the measurement settings $a,b$ made on one side can affect the state of affairs on the other side.
However, we cannot exclude the existence of correlations.
In the r.h.s of (\ref{eq:lawofp}), we can have that
\begin{eqnarray}
P(A\mid B,a,b) &\neq& P(A\mid a)\label{eq:lc_cor1}\\
P(B\mid a,b)   &\neq& P(B\mid b)\label{eq:lc_cor2}
\end{eqnarray}
notwithstanding that events $A$ and $a$ are spacelike separated from $B$ and $b$.
However, relativistic causality requires the correlations implied by (\ref{eq:lc_cor1}) and (\ref{eq:lc_cor2}) to be explained by local common causes $\lambda$.
They are local because they are supposed to lie at the intersection of the backward light cones of the measurement events.
Once the common causes $\lambda$ are specified, the inclusion of spacelike separated parameters in the l.h.s of (\ref{eq:lc_cor1}) and (\ref{eq:lc_cor2}) become redundant
\begin{eqnarray}
P(A\mid B,a,b,\lambda) &=& P(A\mid a,\lambda)\label{eq:lc_cor3}\\
P(B\mid a,b,\lambda)   &=& P(B\mid b,\lambda)\label{eq:lc_cor4}
\end{eqnarray}
Including $\lambda$ in (\ref{eq:lawofp})
\begin{equation}\label{eq:lawofpL}
P(A,B\mid a,b,\lambda)=P(A\mid B,a,b,\lambda)P(B\mid a,b,\lambda)
\end{equation}
Replacing (\ref{eq:lc_cor3}) and (\ref{eq:lc_cor4}) in (\ref{eq:lawofpL})
\begin{equation}\label{eq:lc}
P(A,B\mid a,b,\lambda)=P(A\mid a,\lambda)P(B\mid b,\lambda)
\end{equation}
The last equation is also known as the screening-off condition.
It is the formal expression of the intuitive idea behind relativistic locality and is Bell's definition of LC for the case at hand.

The common cause $\lambda$ is usually called ``hidden variables''; however, it is somewhat misleading to believe the $\lambda$ variables are necessarily unknown parameters.
The only condition they need to comply with is lying at the intersection of the backward light cones of the measuring events to constitute a local explanation of the correlations.
Appendix \ref{app:CCM} further expands on the meaning of the common causes.
\subsubsection{An alleged counterexample disproving LC}\label{app_sssec:Aace}
Although the following example might seem trivial to the cognoscenti, the puzzling nature of quantum entanglement and the reason it would imply nonlocality are sometimes surprisingly misunderstood, even by researchers working in quantum information. 

This problem is related to a lack of awareness about the meaning of local causality, which leads to confusing a common cause with a ``local'' common cause. For instance, the preparation of an entangled pair is surely a common cause realized in its quantum entangled state (3); however, as proved by (5), it fails as a local common cause because it cannot screen off spacelike separated events.

A concrete case of a usual misunderstanding is useful to clarify the LC concept.\footnote{This paper was rejected several times based on arguments of this type, so we think it is worth working out a concrete example.}
The following purports to be a counterexample proving that the local causality condition (\ref{eq:lc}) is incorrect because it fails even when there is an obvious local explanation for exact anti-correlations.

Let us consider an experiment in which pairs of particles are created together.
The setup is such that when one particle is labeled red (R), the other is labeled blue(B), and each color is created with equal probability.
The particle in each pair is then sent to two distant laboratories, L1 and L2, where its color is tested upon arrival.

This game is described by the following statistical distribution,
{\small
\begin{equation}
\left.
\begin{array}{ccccc}
P(R,B\mid L1,L2) &=& P(B,R\mid L1,l2) &=&1/2\\
P(R,R\mid L1,L2) &=& P(B,B\mid L1,l2) &=&0\\
P(R\mid L1)      &=& P(B\mid L1)           &=&1/2\\
P(R\mid L2)      &=& P(B\mid L2)           &=&1/2
\end{array}
\right\}
\end{equation}
}
This trivial game reproduces the singlet quantum probabilities, has an obvious local explanation for its perfect anti-correlations, and, however, it seems to fail to comply with the LC criterion since, for instance,
\begin{equation}\label{eq:sslc_aux1}
\underbrace{P(R,B\mid L1,L2)}_{1/2} \neq \underbrace{P(R\mid L1)*P(B\mid L2)}_{1/2\,*\,1/2\,=\,1/4}
\end{equation}
Note the similarity between (\ref{eq:sslc_aux1}) and (\ref{eq:sslc}), albeit (\ref{eq:sslc}) is supposed to prove nonlocality while (\ref{eq:sslc_aux1}) has an obvious local explanation.

The apparent paradox is resolved by noting that while in (\ref{eq:sslc}) we already included all the information we have about the preparation at the source by putting $\lambda=\mid\psi\rangle$, that information is still missing in (\ref{eq:sslc_aux1}).

For the case of (\ref{eq:sslc_aux1}), the preparation at the source is described by setting $\lambda=(R,B)$ or $\lambda=(B,R)$, each with probability $1/2$ so that,
\begin{equation}
\left.
\begin{array}{ccc}
P(R,B\mid L1,L2, \lambda=(R,B)) &=&1\\
P(B,R\mid L1,L2, \lambda=(R,B)) &=&0\\
P(R\mid L1,\lambda=(R,B))       &=& 1\\
P(B\mid L1,\lambda=(R,B))       &=& 0\\
P(R\mid L2,\lambda=(R,B))       &=& 0\\
P(B\mid L2,\lambda=(R,B))       &=& 1
\end{array}
\right\}
\end{equation}
Similar values correspond when $\lambda=(B,R)$ and it can be easily checked that the local causality criterium (\ref{eq:lc}) is verified in each case, for instance,
{\small
\begin{equation}
\underbrace{P(R,B\mid L1,L2,\lambda=(R,B))}_{=1} = \underbrace{P(R\mid L1,\lambda=(R,B))}_{=1}*\underbrace{P(B\mid L2,\lambda=(R,B))}_{=1}
\end{equation}
\begin{equation}
\underbrace{P(B,R\mid L1,L2,\lambda=(R,B))}_{=0} = \underbrace{P(B\mid L1,\lambda=(R,B))}_{=0}*\underbrace{P(R\mid L2,\lambda=(R,B))}_{=0}
\end{equation}
}
Thus, the lack of a locally causal explanation in the case of the quantum correlations is formally disclosed, and a natural foundational question arises: what is the local explanation for the perfect quantum anti-correlations that prevents the existence of instantaneous influences between the two distant particles?

A common answer asserts that locality is justified because ``realism,'' understood as preexisting values, as in our example where the colors were already ``real'' before the measurement, is false in quantum mechanics.

However, it is erroneous to infer that a local quantum explanation exists because a naive ``local realistic'' explanation is nonexistent. 
It should be obvious that such an inference constitutes a sophism. 
In part, the purpose of this paper is to motivate the analysis of sensible alternatives based on the correct understanding of Bell's and Einstein's reasoning.
\subsection{Common Causes meaning}\label{app:CCM}
Some researchers find it perplexing that the quantum state $\mid\psi\rangle$ can be considered a common cause in the definition of local causality.
That prejudice is mainly owed to three facts:
\begin{itemize}
\item Bell's 1964 theorem was based on the EPR reasoning and a deterministic hidden variable model.
Therefore, it is common to interpret the hidden variables as preexisting elements of physical reality.
\item The former interpretation is reinforced by the inclusion in Bell's 1964 paper of a ``local realistic'' concrete example where $\lambda$ is identified with a preexisting spin vector.
\item Local causality and Reinchenbach's Principle of Common cause are not widely known concepts, consequently so is the fact that LC is not based on determinism and directly applies to quantum mechanics avoiding ontological commitments.
\end{itemize}
Sustaining that $\lambda$, understood as a common cause, is by necessity an element foreign to quantum mechanics amounts to forbidding the application of the local causality concept to quantum mechanics, that according to Bell's reasoning would be unjustified.
The general nature of the $\lambda$ variables and that they may well include the quantum state was explicitly mentioned by Bell\cite{pBel81}:
\begin{quote}
``It is notable that in this argument nothing is said about the locality, or even localizability, of the variable $\lambda$. These variables could well include, for example, quantum mechanical state vectors, which have no particular localization in ordinary space-time.''
\end{quote}
\subsection{Proof of formula (\ref{eq:sslc})}\label{app:FPQNL}
We can test if the quantum state can be a local common cause explaining the perfect anti-correlations by setting $\lambda=\mid\psi\rangle$ in (\ref{eq:lc}) with $\mid \psi\rangle$ given by (\ref{eq:ss}), where $\mid +\rangle$ and $\mid -\rangle$ denote the spin eigenstates in the $z$-direction.
We assume that motion takes place in the $y$ direction with setting angles $a$ and $b$ lying in the $xz$ plane measured with respect to the $z$ axis.
If $\mid a,+\rangle$ and $\mid a,-\rangle$ are the spin eigenstates in the $a$ direction.
\begin{eqnarray}
\mid a,+\rangle &=& +\cos\frac{a}{2}\mid +\rangle + \sin\frac{a}{2}\mid -\rangle\label{eq:amas}\\
\mid a,-\rangle &=& -\sin\frac{a}{2}\mid +\rangle + \cos\frac{a}{2}\mid -\rangle
\end{eqnarray}
Analogously for the particle measured at the other laboratory, we have
\begin{eqnarray}
\mid b,+\rangle &=& +\cos\frac{b}{2}\mid +\rangle + \sin\frac{b}{2}\mid -\rangle\\
\mid b,-\rangle &=& -\sin\frac{b}{2}\mid +\rangle + \cos\frac{b}{2}\mid -\rangle\label{eq:bmenos}
\end{eqnarray}
The joint probability according the quantum formalism is
{\small
\[
P(A,B\mid a,b,\mid \psi\rangle)=\langle\psi\mid (\mid a,A\rangle\otimes\mid b,B\rangle \langle a,A\mid \otimes\langle b,B\mid ) \mid \psi\rangle
\]
}
Letting $A=+1$, $B=-1$ and putting
\begin{eqnarray}
P^*=P(+1,-1\mid a,b,\mid \psi\rangle)
\end{eqnarray}
according to (\ref{eq:ss}), (\ref{eq:amas}) and (\ref{eq:bmenos}),
{\small
\begin{eqnarray}
P^* &=& \langle\psi\mid (\mid a,+\rangle\otimes\mid b,-\rangle \langle a,+\mid \otimes\langle b,-\mid ) \mid \psi\rangle\\
                          &=& \frac{1}{\sqrt{2}}(\langle +\mid a,+\rangle \langle -\mid b,-\rangle -\langle -\mid a,+\rangle \langle +\mid b,-\rangle)\frac{1}{\sqrt{2}}()^*\nonumber\\
                          &=&\frac{1}{2}(\cos\frac{a}{2}\cos\frac{b}{2}+\sin\frac{a}{2}\sin\frac{b}{2})^2\\
                          &=& \frac{1}{2}\cos^2(\frac{a-b}{2})\label{eq:a_b}
\end{eqnarray}
}
Where $()^*$ represents the complex conjugate of the first factor in parenthesis.
If we further assume $a=b$, (\ref{eq:a_b}) gives
\begin{eqnarray}
P^*=P(+1,-1\mid a,a,\mid \psi\rangle) &=& \frac{1}{2}\label{eq:lhd}
\end{eqnarray}
When we perform a measurement only in Alice's laboratory, the quantum formalism prescribes
{\small
\begin{eqnarray}
P(+1,a,\mid \psi\rangle) &=& \langle\psi\mid ( \mid a,+\rangle\langle a,+\mid \otimes I )\mid \psi\rangle\nonumber\\
                     &=& \langle\psi\mid [( \mid a,+\rangle\langle a,+\mid \otimes I )\mid \psi\rangle]\nonumber\\
                     &=& \langle\psi\mid \left[\frac{1}{\sqrt{2}}( \mid a,+\rangle\langle a,+\mid +\rangle\otimes \mid -\rangle -\mid                      a,+\rangle\langle a,+\mid -\rangle\otimes \mid +\rangle)\right]\nonumber\\
                     &=& \langle\psi\mid \left[\frac{1}{\sqrt{2}}( \cos\frac{a}{2}\mid a,+\rangle\otimes \mid -\rangle -\sin\frac{a}{2}\mid a,+\rangle\otimes \mid +\rangle)\right]\nonumber\\
                     &=& \frac{1}{2}\left[\cos\frac{a}{2}\langle +\mid a,+\rangle + \sin\frac{a}{2}\langle -\mid a,+\rangle\right]\nonumber\\
                     &=& \frac{1}{2}\left[\cos^2\frac{a}{2} + \sin^2\frac{a}{2}\right]\nonumber\\
                     &=& \frac{1}{2}\label{eq:rhd1}
\end{eqnarray}
}
Where $I=\mid +\rangle\langle+\mid +\mid -\rangle\langle-\mid $ is the identity operator in the one particle two-dimensional Hilbert-space.
In a similar way, performing a measurement only on Bob's laboratory we find
{\small
\begin{eqnarray}
P(-1,b,\mid \psi\rangle) &=& \langle\psi\mid ( I \otimes \mid b,-\rangle\langle b,-\mid  )\mid \psi\rangle=\frac{1}{2}\label{eq:rhd2}
\end{eqnarray}
}
From (\ref{eq:lhd}), (\ref{eq:rhd1}), and (\ref{eq:rhd2}), we obtain (\ref{eq:sslc}) formally proving that ordinary quantum mechanics lacks a local common cause explanation for its correlations.
\newpage
\bibliography{zBellbibfile}
\end{document}